# Jupiter – friend or foe? I: the asteroids

**J. Horner and B. W. JONES**

*Astronomy Group, Physics & Astronomy, The Open University, Milton Keynes, MK7 6AA, UK*

*e-mail: j.a.horner@open.ac.uk    Phone: +44 1908 653229        Fax: +44 1908 654192*

*Received 12 March 2008, accepted 30 May 2008*

**(SHORT TITLE: Jupiter – friend or foe? I: the asteroids)**

**Abstract**

The asteroids are the major source of potential impactors on the Earth today. It has long been assumed that the giant planet Jupiter acts as a shield, significantly lowering the impact rate on the Earth from both cometary and asteroidal bodies. Such shielding, it is claimed, enabled the development and evolution of life in a collisional environment which is not overly hostile. The reduced frequency of impacts, and of related mass extinctions, would have allowed life the time to thrive, where it would otherwise have been suppressed. However, in the past, little work has been carried out to examine the validity of this idea. In the first of several papers, we examine the degree to which the impact risk resulting from a population representative of the asteroids is enhanced or lessened by the presence of a giant planet, in an attempt to fully understand the impact regime under which life on Earth has developed. Our results show that the situation is far less clear cut that has previously been assumed – for example, the presence of a giant planet can act to *enhance* significantly the impact rate of asteroids at the Earth.

**Introduction**

Throughout Earth history, our planet has suffered impacts from asteroidal and cometary material. As well as disrupting the landscape, the larger of these impacts have had effects that led to climate changes, usually short-lived, that in turn have led to the extinction of a large proportion of species in the biosphere (Morris (1998)).

Anyone who has watched popular science programmes which discuss the effect of impacts on the Earth, along with their implications for the survival of life, will have encountered the idea that Jupiter acts to lower significantly the flux of objects that hit the Earth. The inference, sometimes explicitly stated, is that Jupiter's role in the evolution of life on our planet is surprisingly large (see, for example, Ward and Brownlee (2000), http://tinyurl.com/2g59ee, http://tinyurl.com/2yvk6x, and http://tinyurl.com/34msr8, for examples of the pervasiveness of this idea). It is claimed that, in preventing the great majority of threatening objects from encountering the Earth, Jupiter has significantly lowered the rate at which impact-driven mass extinctions happen, giving life time to get a foothold, and then evolve to its great present day diversity. Were Jupiter absent, so it is claimed, then the Earth could have suffered large impacts so frequently that it might not have acquired advanced life, or even be barren.

These arguments are quite widely accepted in the academic world, but when one looks back through the literature, it seems that, until recently, very little work has been carried out to examine in detail the effects of the giant planet on the flux of cometary and asteroidal bodies through the inner Solar System. It has been suggested that, in systems containing only "failed Jupiters" (bodies which grew to the size of, say, Uranus and Neptune, but failed to develop beyond that stage), the impact flux experienced by any terrestrial planets would be a factor of a thousand greater than that seen in our system today (Wetherill, 1994). This is because of the less efficient ejection of material from the Solar System during its early days. However, very little work exists to support or argue against this conclusion, and it is unclear how it would be affected by the current understanding of planet formation.

Laakso et al. (2006) approached the question from a different angle. Using numerical integration, they examined the effect of the position and mass of a Jovian planet on the rate of ejection of particles placed on eccentric orbits that initially crossed the habitable zone (being the range of distances from a star within which water at the surface of an "Earth" would be stable in the liquid phase, liquid water being essential for all forms of life on Earth). They used our Solar System as a test case for their method, and found the surprising result that Jupiter "*in its current orbit, may provide a minimal amount of protection to the Earth*". Despite this, the idea that "Jupiters" automatically lower the impact rate in planetary systems is well entrenched in astronomical thinking, and the lack of planets analogous to Jupiter has been used to explain observations such as that of a significant dust excess around the star Tau Ceti (Greaves, 2006). However, questions about Jupiter's effect on the terrestrial impact record have been raised in relation to the Late Heavy Bombardment. If the "Nice model" (Gomes et al., 2005) is considered, for example, then it is clear that removing Jupiter from our Solar System would greatly lessen or remove the effects of the Late Heavy Bombardment on our young planet.

In our opinion, it seems that the idea of "Jupiter, the protector" dates back to the days when the main impact risk to the Earth was thought to arise from the population of long period comets (LPCs), falling inwards from the Oort cloud. The majority of such objects are expelled from the Solar System on their very first pass as a result of Jovian perturbations, hence lowering the chance of one of these cosmic bullets striking the Earth.

In recent times, however, it has been estimated that among the near Earth objects (NEOs) i.e. asteroids and comets that make close approaches to the Earth, the comets contribute only a few percent of the population (Bottke et al. (2002), Chapman and Morrison (1994)). Among the comets most are short period comets (SPCs), so the LPCs contribute only slightly to the NEOs. (Near the Earth, comets generally move much faster that asteroids, and so the effect of an impact of a body of given mass, will be greater for a comet.)

For the NEOs, the role of Jupiter as friend or foe is far less clear than in the case of the LPCs alone, as can be demonstrated by a thought experiment. Imagine our Solar System as it is today, and remove Jupiter entirely. At one fell swoop, you have removed the main driving force which transfers asteroidal bodies from the main belt between Mars and Jupiter (where the great majority lie) to the inner Solar System. Furthermore, you have lost the object which is the dynamical source and controller of the great majority of the short period comets (SPCs). On the other hand, you have also lost the object most efficient at removing debris from the inner Solar System, though if the detritus is not being put there in anywhere near the same quantity as with Jupiter present, then removal is less important.

Overall, the situation is no longer clear cut. What Jupiter gives with one hand, it may take away with the other. In order to study the exact relationship between the giant planet and the impact rate on the Earth, we decided to run a series of *n*-body simulations to see how varying the mass of a giant planet in Jupiter's orbit would change the impact rate on Earth. Since there are three distinct populations which provide the main impact threat to the Earth (the asteroids (sourced from the Main Belt (Morbidelli et al. (2002)), the SPCs (which come from the trans-Neptunian region, Horner and Evans (2006)), and the LPCs (which come from the Oort Cloud, Oort (1950)), we decided to split the problem three ways, and examine each population in turn. This paper details our results for the asteroids, an entirely different reservoir of bodies to that studied by Wetherill (who studied the LPCs), and generally accepted to be the most important population of potential Earth impactors. The SPC and the LPC components of the impact risk will be detailed in later work.

Whereas the work here advances our understanding of what jovian characteristics have determined the bombardment suffered by the Earth, it also advances our understanding of the requirements for the habitability of "Earths" in exoplanetary systems.

**Simulating the impact flux**

Of the three parent populations that supply Earth impacting bodies, the most copious is the asteroids. However, in creating a swarm of test asteroids which might evolve on to Earth impacting orbits, we face huge uncertainties, particularly relating to $N(a)$ at the start of a simulation ($t = 0$), where $N(a)$ is the number of asteroidal bodies as a function of semimajor axis $a$. That Jupiter has been perturbing the orbits of the objects currently observed in the asteroid belt in our own Solar System since its formation means that using the current belt as the source would be misguided. It is therefore important to attempt to construct a far less perturbed initial population for the asteroid belt, if one wishes to observe the effect of changing Jupiter's mass on the impact rate. However, this is more easily said than done. In Appendix 1, we discuss in some detail how we constructed such a test population for use in this work. We finally settled on a population distribution, $N(a)$ at $t = 0$ given by (see Appendix 1)

$$N_0(a) = k(a - a_{min})^{1/2} \tag{1}$$

where $k$ is a constant and $a_{min}$ is the inner boundary of the asteroid distribution. The value of $a_{min}$ was chosen to be 1.558 AU, equivalent to the orbital semi-major axis of the planet Mars[i], plus three Hill radii, while the outer boundary, $a_{max}$, was placed three Hill radii within the orbit of the giant planet i.e. interior to the 5.202 AU of Jupiter's orbit. (To read footnotes, hover over the number.) Closer to the planets than these two distances, asteroidal bodies are unlikely to form. (The Hill radius gives the distance between two bodies, such as a planet and another body, at which their gravitational interaction is of the same order as the gravitational interaction of each body with the star they orbit. Three Hill radii of a planet is its "gravitational reach". It is given by $R_H = a_p \left( \frac{M_{planet}}{3 M_{Sun}} \right)^{1/3}$ where $M$ is mass.) It is important to note that our main conclusions below concerning the variations of the impact rate on Earth as a function of giant planet mass are not sensitive to the precise form of $N_0(a)$. The placement of the inner and outer edges at $3R_H$ beyond the orbit of the planets in question was chosen as a reasonable compromise between placing the edge of the belts far enough away from the planet so as not to experience significant perturbations in the early stages of the simulations, and placing the edge so distant that the belt itself would be unfairly constrained.

To generate the values of $a$ for our population of asteroids the cumulative probability distribution corresponding to $N_0(a)$ was sampled by a random number generator to generate $10^5$ values of $a$ between $a_{min}$ and $a_{max}$. The other five orbital elements for each asteroid were randomly allocated, as follows. The orbital inclination, $i$, was randomly sampled from the range 0-10°, and the eccentricity, $e$, randomly allocated from the range 0.0-0.10. These ranges encompass the majority of the known asteroids today. In the distant past, at the start of our simulations, an even greater proportion would have been encompassed. They represent a disk of solid material that has received a moderate, but not excessive, amount of stirring during the formation of the planets (e,g, Ward 2002). The remaining three orbital elements – the longitude of the ascending node, the argument of perihelion, and the mean anomaly, were each randomly selected from the range 0-360°. (For a brief description of orbital elements, see, for example, Jones 2007(a).)

We simulated these orbits for a period of 10 million years using the hybrid integrator contained within the *MERCURY* package (Chambers, 1999), along with the planets Earth, Mars, Jupiter, Saturn, Uranus and Neptune. We take $t = 0$ to be the moment when Jupiter became fully formed. The integration duration was chosen to provide a balance between the required computation time and the statistical significance of the results obtained. The Earth within our simulations was artificially physically inflated to have a radius of 1 million kilometres, in order to enhance the impact rate from objects on Earth crossing orbits. Simple initial integrations were carried out to confirm that this inflation did affect the impact rate as expected, with the rate scaling with the cross-sectional area of the planet (the effect of gravitational focussing on the impact rate was observed to be negligible). The asteroidal bodies interact gravitationally with the Sun and planets, but not with each other – they are treated as massless which is a good model as a typical asteroidal body is normally at least $10^{11}$ times less massive than Jupiter!

The "Jupiter" used in our runs was modified so that we ran 12 separate masses. In multiples of Jupiter's mass $M_J$ these are: 0.01, 0.05, 0.10, 0.15, 0.20, 0.25, 0.33, 0.50, 0.75, 1.00, 1.50, and 2.00. Hereafter, we refer to these runs by the mass of the planet used, so that $M_{1.00}$ refers to the run using a planet of 1.00 $M_J$, and $M_{0.01}$ refers to the run using a planet of 0.01 $M_J$, and so on. The orbital elements for the "Jupiter" were identical in all cases to those of Jupiter today. Similarly, the elements taken for the other planets in the simulations were identical to those today – the only difference in the planetary setup between one run and the next was the change in Jovian mass – all other planetary variables were held constant.

It is obvious that, in reality, were Jupiter a different mass, the architecture of the outer Solar System would likely be somewhat different. However, rather than try to quantify the uncertain effects of a change to the formation of our own Solar System, we felt it best to change solely the mass of the "Jupiter", and therefore work with a known, albeit modified, system rather than an uncertain theoretical construct. In the case of the flux of objects moving inwards from the asteroid belt, this does not seem a particularly troublesome assumption, because Jupiter is by far the dominant influence on the asteroids.

The complete suite of integrations ran for some six months of real time, spread over the cluster of computers sited at the Open University. This six months of real time equates to over twenty years of computation time, and resulted in measures of the impact flux for each of the twelve "Jupiters". The eventual fate of each asteroidal body was also noted.

**Results**

In this section, we present the results of our simulations, leaving discussion for the next section. This should allow the reader to be familiar with the results, and perhaps reach their own conclusions, before we present a detailed discussion.

Figures 1 to 3 show a variety of different results from our simulations (these Figures are at the end of the article). Figure 1 shows the evolution of our test populations as a function of time for $M_{0.25}$ (Figure 1a) and $M_{1.00}$ (Figure 1b). Five temporal snapshots are shown, detailing the distribution of asteroidal bodies at $t = 0$ Myr (the start of the simulation), 1 Myr, 2 Myr, 5 Myr and 10 Myr (the end of our simulation). In order to give a fair representation of the asteroid distributions, Figure 1 shows the number of objects located in rings of equal width (in semimajor axis), working outward from a semimajor axis of 1.5 AU to 5.5 AU. This space is broken up into 1000 equal width bins, so that the width of each bin is 0.004 AU. In effect, this means that the maximum initial population in any bin is less than 400 objects, and so the $y$-axis in all of the plots in Figure 1, extends from 0 to 400. The initial populations were, as described above, distributed according to equation 1, with inner and outer limits fixed as described. Note that the location of the outer edge of the belt changes between the two plots, in response to the larger Hill sphere of the more massive Jupiter in Figure 1b. Additionally, it is clear that the initial population is somewhat scattered, a result of the random number generator used to select $a$. As an aid for the reader, the points corresponding to each bin have been connected, which makes small details easier to see. The development of fine structure in the belts is clearly apparent as early as 1 Myr, and this structure continues to develop through the period of the simulations. Equivalent plots for all 12 "Jupiter" masses can be found in Appendix 2.

> Figure 1  The evolution of the asteroid populations as a function of time: (a) variation of population at $M_{0.25}$ (b) variation of population at $M_{1.00}$

Figure 1(a) shows the behaviour of asteroids in the case where the "Jupiter" has a mass 0.25 $M_J$, while Figure 1(b) shows the evolution of the objects in the case where the "Jupiter" has the same mass as ours ($M_{0.25}$ and $M_{1.00}$ cases respectively). The five time slices shown in each plot are, from top to bottom, $t = 0$, 1, 2, 5 and 10 Myr (the end of the simulations). Equivalent plots are given in Appendix 2 for all 12 "Jupiter" masses. The y-axis extends to about 400 in both cases, but is a function of bin width.

Figure 2 shows the final populations (at 10 Myr) in the $M_{0.25}$ and $M_{1.00}$ cases. In order to allow easy comparison, the $M_{0.25}$ results have been inverted, and placed below those for the $M_{1.00}$ case. A number of differences are striking, and will be discussed in some detail. Between the two distributions, a number of + marks show the location of various of the Jovian mean motion resonances[ii]. Working from left to right, the resonances shown are 1:6, 1:5, 1:4, 2:7, 1:3, 3:8, 2:5, 3:7, 1:2, 4:7, 3:5, 5:8, 2:3, 5:7, 3:4, 4:5 and 1:1. It is clear that these resonances play an important role in the evolution of asteroid belts, and we will discuss them further in the following section.

> Figure 2  The final asteroid distributions for the two cases $M_{0.25}$ (inverted, lower) and $M_{1.00}$ (upper). This Figure allows the direct comparison of the final distributions between these two sample cases. Between the two distributions, a series of + marks show the location of a number of key mean motion resonances with the "Jupiter". From left to right, the resonances shown are 1:6, 1:5, 1:4, 2:7, 1:3, 3:8, 2:5, 3:7, 1:2, 4:7, 3:5, 5:8, 2:3, 5:7, 3:4, 4:5 and 1:1. For a more detailed explanation of resonances, see the discussion section.

Figure 3 shows the evolution with time of the number of collisions of asteroidal bodies with the inflated Earth as a function of "Jupiter" mass. The lines, in ascending order from the x-axis, show the total number of collisions versus mass that had occurred at 1, 2, 5 and 10 Myr. The form of these graphs will be discussed in detail in the next section, but note that the final two time slices (5 and 10 Myr) show that the *form* of the graphs has settled down.

> Figure 3  Plot showing the number of collisions with the inflated Earth as a function of "Jupiter" mass. The curves show the total number of collisions at a variety of times. Working upwards from the x-axis, the times are 1, 2, 5 and 10 Myr. The total numbers at 10 Myr are presented in Table 1.

**Discussion**

Figure 3, which illustrates our core result, is discussed first, then Figures 1 and 2.

From Figure 3 it is clear that the notion that any "Jupiter" would provide more shielding than no "Jupiter" at all is incorrect, at least for impactors originating from the asteroid belt. It seems that the effect of a "Jupiter" on the impact flux on potentially habitable worlds is far more complex than was initially thought. With our current Jupiter ($M = 1.0\ M_J$), potentially impacting objects seem to be ejected from the Solar System with such rapidity that they pose rather little risk for planets in the habitable zone (such as the Earth), and therefore, Jupiter offers a large degree of shielding, compared to "Jupiters" of smaller mass, down to about 0.1 $M_J$. You can see from Figure 3 that planets more massive than Jupiter offer little further improvement.

At the other end of the scale, at very small "Jupiter" masses, fewer asteroidal objects are scattered onto orbits which cross the habitable zone, and so, once again, the impact rate is low. The more interesting and complicated situation occurs for intermediate masses, where the giant planet is massive enough to emplace asteroidal objects on threatening orbits, but small enough that ejection events are still infrequent. The situation which offers the greatest enhancement to the impact rate is one located around 0.20 $M_J$, in our simulations, at which point the planet is massive enough to efficiently inject objects to Earth-crossing orbits, but small enough that the time spent on these orbits is such that the impact rate is significantly enhanced. Had we used a different form of $N_0(a)$, the peak could well have been at a different intermediate mass (due to the shifting concentration of material in areas swept by secular resonances), but the broad picture in Figure 3 would be the same. (The double peak in Figure 3 is not a large feature and is possibly a statistical fluctuation, though time consuming further study would be needed to investigate whether this is so.)

The effects of the other planets, particularly Saturn and Mars, are pretty much constant between the different runs. However, due to the reduction in the Jovian effect at the lower "Jupiter" masses (particularly below $M < 0.2 M_J$), these planets play a more significant role in these cases, relatively, than at higher "Jupiter" masses, as the overwhelming and masking effects of the "Jupiter" are removed, allowing the effects of the smaller planets to be more clearly observed, and giving them longer to act.

From this we can see that our Jupiter is approximately as effective a shield as a giant planet of about 0.05 $M_J$, which is 15.9 Earth masses (c.f. 14.5 and 17.1 Earth masses for Uranus and Neptune respectively). The $M_{0.01}$ point (0.01 $M_J$) corresponds to a planet with a mass just 3.18 times that of the Earth. It is possible that in this case a planet would form in the asteroid region in the order of 10 million years, much depleting the asteroid population (e.g.Wetherill (1991)). In this case the reduction in asteroid numbers could well, in the long term, reduce the number of collisions subsequent to the planet's formation below that at 10 Myr in Figure 3 (as a result of the planet acting to clear its immediate vicinity through the accretion and ejection of material). Clearly, this planetary system would be significantly different to our own. The discussion of such hypothetical systems is beyond the scope of this work (though we intend to study the complicated problem of alien planetary systems in future work).

Let's turn now to the number of collisions as a function of time. Figure 1 shows a rapid emergence of structures as time passes. In the $M_{0.25}$ case (Figure 1(a)) the most obvious features are the depletion of asteroids in the outer area of the asteroid belt, and the sharp "spiky" distribution in this region, a result of the effect of mean motion resonances (MMRs), and a large depleted area around 2.5 AU, which is the result of strong secular resonances[iii] involving Jupiter. These are discussed in more detail shortly. In the case of $M_{1.00}$ (Figure 1(b)), a variety of similar features are visible. In fact, at first glance, the distributions appear strikingly similar. However, on closer inspection, a number of significant differences become apparent.

First, in the $M_{1.00}$ case, the asteroid belt is truncated at a smaller heliocentric distance (~4.0 AU vs. ~4.5 AU – it should be noted that in both cases this outer edge has been trimmed to be somewhat closer to the Sun than that of the initial population). Second, the severe depletion around 2.5 AU has shifted to just beyond 2 AU. This is evidence of how the location of the secular resonances in the asteroid belt is a function of the mass of the Jovian planet, whereas the locations of the MMRs are purely determined by the location of that planet alone (though the widths of these resonances, and their strengths *are* affected by the planet). In passing, we should note that it is well known that MMRs can cause depopulation, as at 3.28 AU (the 1:2 resonance) in both Figures, or help to enhance the population, as can be seen from the small "spikes" located at the orbit of the giant planet (the 1:1 resonance at 5.2 AU, showing objects captured as Jovian Trojans), again in both Figures. The latter corresponds to the temporary capture of objects in Jupiter-like orbits, in a manner similar to that shown for the Centaurs, the parent population of the SPCs (Horner & Evans 2006).

Figure 2 allows the reader a better opportunity to see the detailed effects of MMRs on the belt. The + symbols mark the locations of a variety of such resonances (as detailed in the Figure caption), and it is clear that they have played an important role in shaping the young asteroid belts. Note again the 1:2 MMR at 3.28 AU clearly leading to depletion in both the $M_{1.00}$ and $M_{0.25}$ belts. What is also clearly visible with this resonance is the way that, as the mass of the "Jupiter" increases, the width of the MMR also increases – this is the case for all MMRs.

On the other hand, the effects of secular resonances as the belt evolves show a different variation as a function of planetary mass. The *location* of these resonances moves with changing Jovian mass, and so they effectively "sweep" through the belt as the mass of the planet increases. It is well known that a resonance called $\nu_6$ marks the inner edge of the asteroid belt in our Solar System (the effect of this resonance can be seen in Figure 2 at around 2 AU in the $M_{1.00}$ case). However, at lower Jovian masses, this resonance lies well within the belt, and results in a broad area of instability (clearly visible at around 2.5 AU in the $M_{0.25}$ plot), which is probably the main route by which lower "Jovian" masses lead to enhanced impact fluxes.

It is clear, therefore, from the examination of Figures 1 and 2, together with those shown in the Appendix 2, that the effects of secular and mean motion resonances play an important role in the removal of objects from the asteroid belt. While the MMRs are locked in semimajor axis, as the mass of the planet is increased, the secular resonances "sweep" through the belt, bringing instability to areas which would otherwise be stable on long timescales. This is clear from Figure 1 and the Figures in Appendix 2 – as the mass of "Jupiter" increases, a secular resonance steadily moves towards the inner edge of the asteroid belt. It also deepens and widens. This doubtless plays a major role in the size and variation of the impact flux on a terrestrial world in these simulations. In fact, it seems quite likely that the evolution of this resonance is the biggest single factor in the rise and fall of the impact rate visible in Figure 3. We believe it to be the $\nu_6$ resonance. Further study of this resonance (and perhaps others) in relation to our data is needed, but it will be time consuming.

However, a detailed discussion of resonant behaviour is beyond the scope of this work, and indeed, such behaviour is already very well explained in the literature (e.g. Murray & Dermott 1999(a) and (b)), so we will leave our discussion of such resonant effects here.

In Table 1, we present the numerical results of our twelve sets of simulations. The various columns detail the mass of the Jovian planet used, the number of impacts (collisions) experienced by the inflated Earth, the number of objects which impact other bodies in the Solar System, the number ejected (in our simulations, any object which reached a heliocentric distance of 1000 AU was considered ejected, and was removed from the calculations), and the number which remain somewhere within the Solar System at the end of the 10 Myr simulations. The variation in the number of objects ejected and remaining in the simulations is far lower than the variation in the impact rate on the Earth. In fact, the simulation in which the fewest asteroids survived is also that which showed the most impacts on the Earth – further evidence of the hugely destabilising effect of the planet in this case. It is interesting to note how the various bodies in our simulations fared, as a whole. Summed over the 12 different setups, the Earth was hit 157794 times (a result of its inflated size), while Mars received 1271 impacts, Jupiter 7783, Saturn 3424, Uranus 32 and Neptune 20. The Sun was hit a total of 558 times, although it should be stressed that, due to the time step chosen for our integrations, we would expect objects dropping to such low perihelion distances to be poorly dealt with in the integrator, so this number should be taken with a large pinch of salt! The effect of inflating the Earth is clearly visible, and given the small numbers of impacts on other bodies, fully justified.

Table 1 The fate of asteroidal bodies. At $t = 0$ there are $10^5$ bodies.
The figures in the table are asteroid numbers $n$ at $t = 10$ Myr.

| $M$(Jupiter masses) | $n_{\text{Earth-impact}}$ | $N_{\text{other impact}}$ | $N_{\text{ejected}}$ | $n_{\text{remaining}}$ |
|---|---|---|---|---|
| 0.01 | 2930 | 1730 | 11166 | 84174 |
| 0.05 | 10875 | 1612 | 16104 | 71409 |
| 0.10 | 18107 | 1175 | 14083 | 66635 |
| 0.15 | 19642 | 1109 | 13189 | 66060 |
| 0.20 | 17632 | 986 | 13753 | 67629 |
| 0.25 | 18294 | 915 | 14206 | 66675 |
| 0.33 | 16063 | 926 | 15611 | 67400 |
| 0.50 | 13560 | 937 | 16746 | 68757 |
| 0.75 | 11447 | 986 | 18088 | 69479 |
| 1.00 | 10233 | 935 | 16897 | 71935 |
| 1.50 | 9841 | 838 | 15316 | 74005 |
| 2.00 | 9169 | 930 | 18413 | 71488 |

The evolution of the various asteroid belts considered above would doubtless continue beyond the end of our short simulations. Indeed, it is likely that the stirring of the belts due to mean motion and secular resonances would continue, and that the belts would slowly shed their less stable members. One factor which would prevent the belts studied from eventually evolving into analogues of that in our own Solar System, even in the $M_{1.00}$ case, is that we do not take account of inter-asteroid interactions in this work (both collisional and gravitational). Nor do we take account of any non-gravitational perturbations, such as the Poynting-Robertson and Yarkovsky effects (Jones 2007(b) and (c) respectively). To incorporate all these features, and to run for the age of our Solar System, presents a huge and daunting technical challenge, and is far beyond the scope of this work. In the future, once computing power has developed enough to handle huge numbers of massive particles in a fully physical environment, such studies will doubtless be feasible and fascinating, but at the moment the incorporation of these features would mean that our simulations would have taken many orders of magnitude longer to run.

**Conclusions**

The idea that the planet Jupiter has acted as an impact shield through the Earth's history is one that is entrenched in standard scientific canon. However, when one looks beyond the general understanding of the impact flux on the Earth, it is clear that little work has been done to examine this idea. In the first of an ongoing series of studies, we have examined the question of Jovian shielding using a test population of particles on orbits representative of the asteroids, one of three reservoirs of potentially hazardous objects, the other two being the SPCs and the LPCs.

The surprising result of this work is that the status of Jupiter as a shield is now under serious question. For an asteroidal population, it seems that our Jupiter is no better as a shield than a far less massive giant planet would be, were it placed on a similar orbit, and that intermediate mass giants enhance the number of collisions. Figure 3 shows that at intermediate mass the number of collisions at 5Myr and 10 Myr is about double that for our Jupiter. If the Earth had suffered double its actual impact rate there would doubtless have been more mass extinctions, though with what outcome for the biosphere today we can only speculate. Certainly, the risk of an impact large enough to wipe all plants and animals from the globe can only increase as the number of impacts increases.

Figures 1, 2, and those in Appendix 2, show that mean motion resonances and at least one secular resonance sculpt the asteroid distribution and are thus responsible for sending impactors our way. We believe that the $\nu_6$ secular resonance plays a major role.

Future work will continue the study of the role of Jupiter in limiting or enhancing the impact rate on the Earth by examining populations of bodies representative of the Centaurs and Trans-Neptunian objects (source of almost all of the SPCs) and the Oort cloud (source of the LPCs, and the population of potential impactors studied by Wetherill in 1994). We will also examine the effect of Jovian *location* on the impact fluxes engendered by the three populations, once studies of the effect of its mass are completed. Given the surprising outcome of the present work we hesitate to anticipate future results, though our integrations of the SPCs (which will follow in paper II) do show a comparable outcome to the work described here.

Additionally, future work will also consider whether the absence of a Jupiter-like body would change the populations of objects which reside in the three reservoirs, a possible effect ignored in this work. Further into the future, we intend to study wholly different planetary systems, using both hypothetical versions of our youthful Solar System and other planetary systems based upon the rapidly expanding field of known exoplanets. The long term goal is to finally answer, once and for all, the question "Jupiter – friend or foe?".

**Acknowledgements**

This work was carried out with funding from PPARC, and JH and BWJ gratefully acknowledge the financial support given by that body.

**References**

Bottke, W.F. et al. (2002). Debiased orbital and absolute magnitude distribution of the near-Earth objects. *Icarus* **156**, 399-433.
Chambers, J.E. (1999). A hybrid symplectic integrator that permits close encounters between massive bodies. *MNRAS* **304**, 793-799.
Chapman, C.R., Morrison, D. (1994). Impacts on the Earth by asteroids and comets: assessing the hazard. *Nature* **367**, 33-40.
Davis, S.S. (2005). The surface density distribution in the solar nebula. *ApJ* **627**, L153-L155
Gomes, R., Levison, H.F., Tsiganis, K., Morbidelli, A. (2005). Origin of the cataclysmic Late Heavy Bombardment period of the terrestrial planets. *Nature* **435**, 466-469.
Greaves, J.S. (2006). Persistent hazardous environments around stars older than the Sun. *International Journal of Astrobiology* **5**, 187-190.
Horner, J. and Evans, N. W. (2006). The capture of Centaurs as Trojans. *Monthly Notices of the Royal Astronomical Society,* **367,** 1, L20-L23.
Jones, B.W. (2007). *Discovering the Solar System, 2$^{nd}$ edition.* John Wiley & Sons, (a) Chapter 1 (b) p79 (c) p84.
Laakso, T., Rantala, J., Kaasalainen, M. (2006). Gravitational scattering by giant planets. *A&A* **456**, 373-378.
Morbidelli, A., Bottke, W.F., Froeschlé, Ch., Michel, P. (2002). Origin and Evolution of Near-Earth Objects. *Asteroids* **III**, 409-422. University of Arizona Press.
Morris, S.C. (1998). The evolution of diversity in ancient ecosystems: a review. *Philosophical Transactions of the Royal Society of London B* **353**, 327-345.
Murray, C.D., Dermott, S.F. (1999). *Solar System Dynamics.* CUP, (a) Chapter 8 (b) Chapter 9.

Oort, J.H. (1950). The structure of the cloud of comets surrounding the Solar System, and a hypothesis concerning its origin. *Bulletin of the Astronomical Institutes of the Netherlands* **XI**, 408, 91-110.

Ward, W.R. (2005). Early clearing of the asteroid belt. *Abstracts of the 36$^{th}$ Lunar and Planetary Science*. Abstract 1491, 2 pages.

Ward, W.R. and Brownlee, D. (2000). *Rare Earth:Why Complex Life is Uncommon in the Universe*. Copernicus, 238-239.

Wetherill, G.W. (1991). Occurrence of Earth-like bodies in planetary systems. Science 253, 535-538.

Wetherill, G.W. (1994). Possible consequences of absence of Jupiters in planetary systems. *Astrophysics & Space Science* **212**, 23-32.

**Appendix 1: The Asteroid Distribution**

In choosing the form of $N_0(a)$, the number of asteroidal bodies per unit interval of semimajor axis $a$ at zero time in our simulations, we faced huge uncertainties. There is a range of models representing the distribution over $a$ of dust and small bodies when the Solar System was young (e.g. Davis 2005). Another uncertainty is to what extent the abundant icy-rocky bodies that formed in the cooler conditions beyond the outer edge of the asteroid belt, mixed inwards.

Whatever the details, it is crucial to remember that gravitational stirring by a giant planet orbiting beyond the asteroid belt prevented the formation of a planet between it and Mars. The asteroidal population must have been largely confined to this zone.

We have used a form for $N_0(a)$ that is similar to the form implicitly favoured by Davis (2005), who derives the surface density of the early solar nebula by a cumulative mass model involving all the planets as they are today. Over the space between Mars and Jupiter

$$N_0(a) = k(a - a_{min})^{1/2} \qquad (1)$$

fits his graph well enough, given the uncertainties. The value of $a_{min}$ has been set by us at three Martian Hill radii beyond the orbit of Mars, and thus at $a_{min}$, $N_0(a) = 0$. This is reasonable because Mars would have cleared bodies closer to its orbit than this, and the asteroid-asteroid collision speeds near $a_{min}$ would have been high, resulting in further depletion. The outer boundary $a_{max}$ at $t = 0$ is at three giant Hill radii interior to the giants orbit. The $(a - a_{min})^{1/2}$ dependence is within the range of possibilities, and gives us a greater number of asteroidal bodies at larger $a$ than some other possible dependences. This is to our advantage because, with the giant planet being far more important than Mars at sending asteroids towards the Earth, it increases the number of collisions for a given $t = 0$ population.

Remember that we are interested in the effect of the mass of the giant planet on the impact rate of asteroidal bodies on the Earth. The exact form of $N_0(a)$ is unlikely to affect our conclusion that Jupiter is no better as a shield than a far less massive giant planet, and that intermediate mass giants are poor shields.

**Appendix 2: Evolution of the various asteroid belts with time.**

The following Figures show the evolution of the asteroid belts as a function of time for each of our 12 "Jupiter" simulations. In order, we show the cases from $M_{0.01}$ to $M_{2.00}$, sequentially by increasing mass. The five time slices shown are take at $t$ = 0 Myr, 1 Myr, 2 Myr, 5 Myr and 10 Myr. The variations in the populations due to the changes in the mass of the giant planet are clear to see. The *x*-axis extends to about 400 in all cases, but is a function of bin width.

$M_{0.01}$

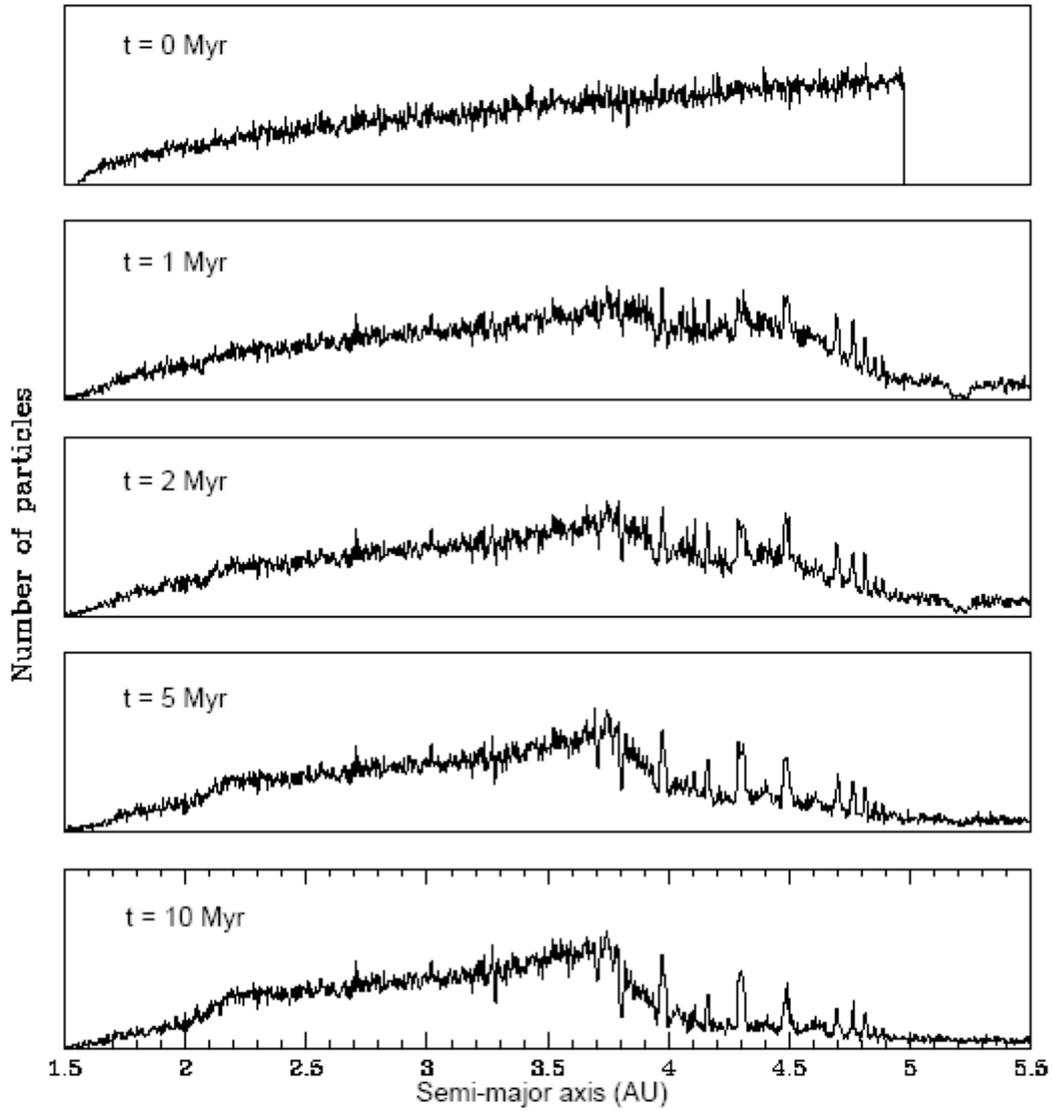

**M**$_{0.05}$

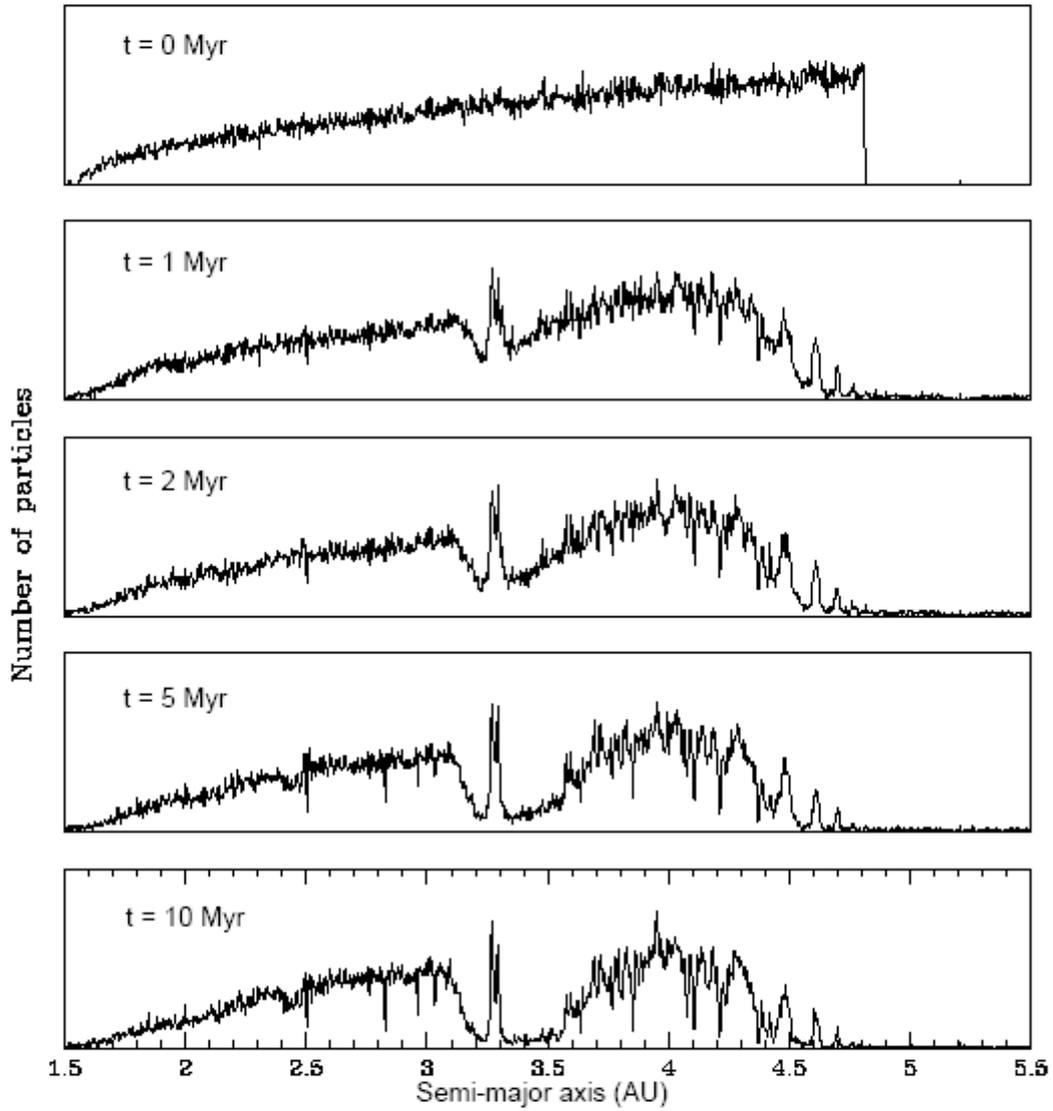

$M_{0.10}$

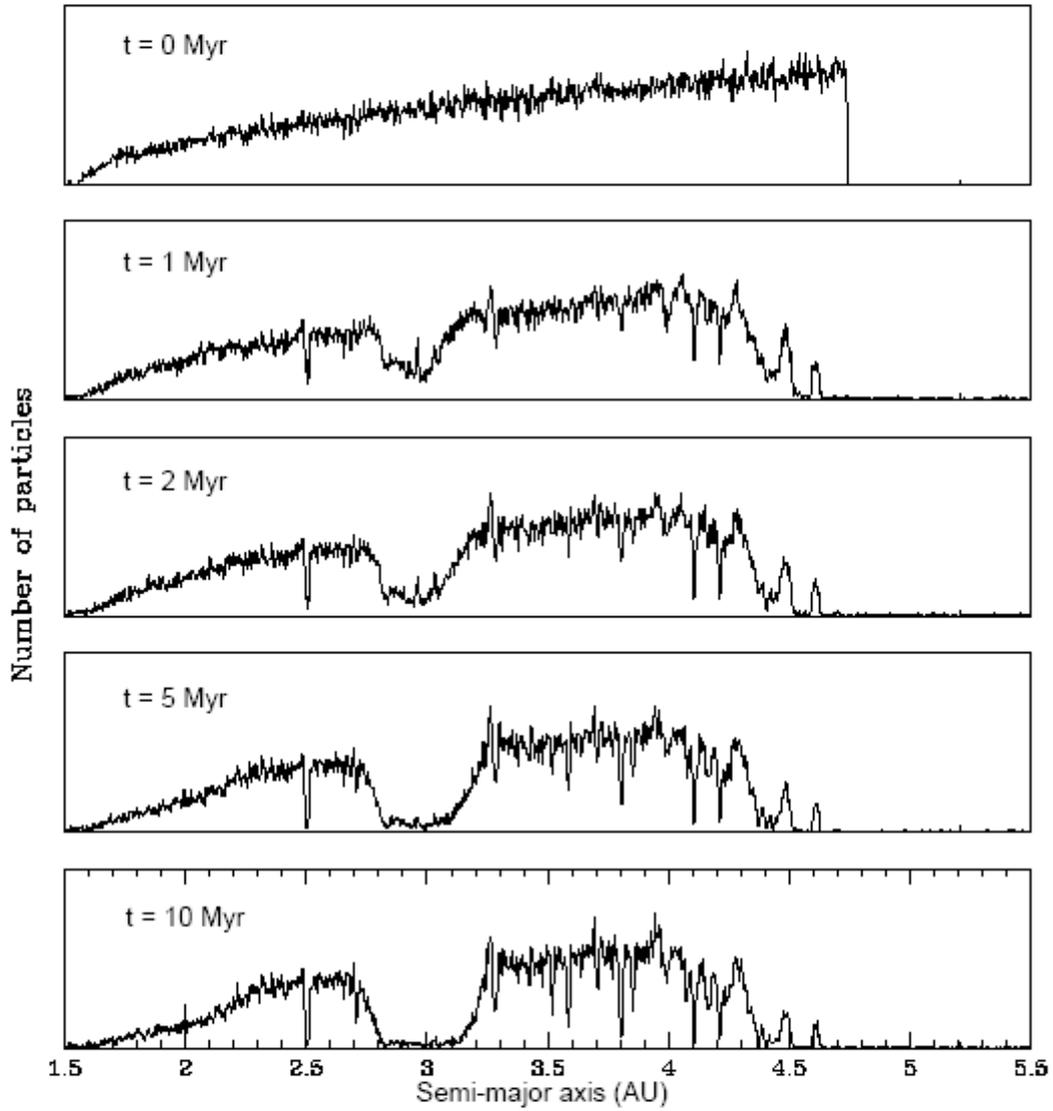

$M_{0.15}$

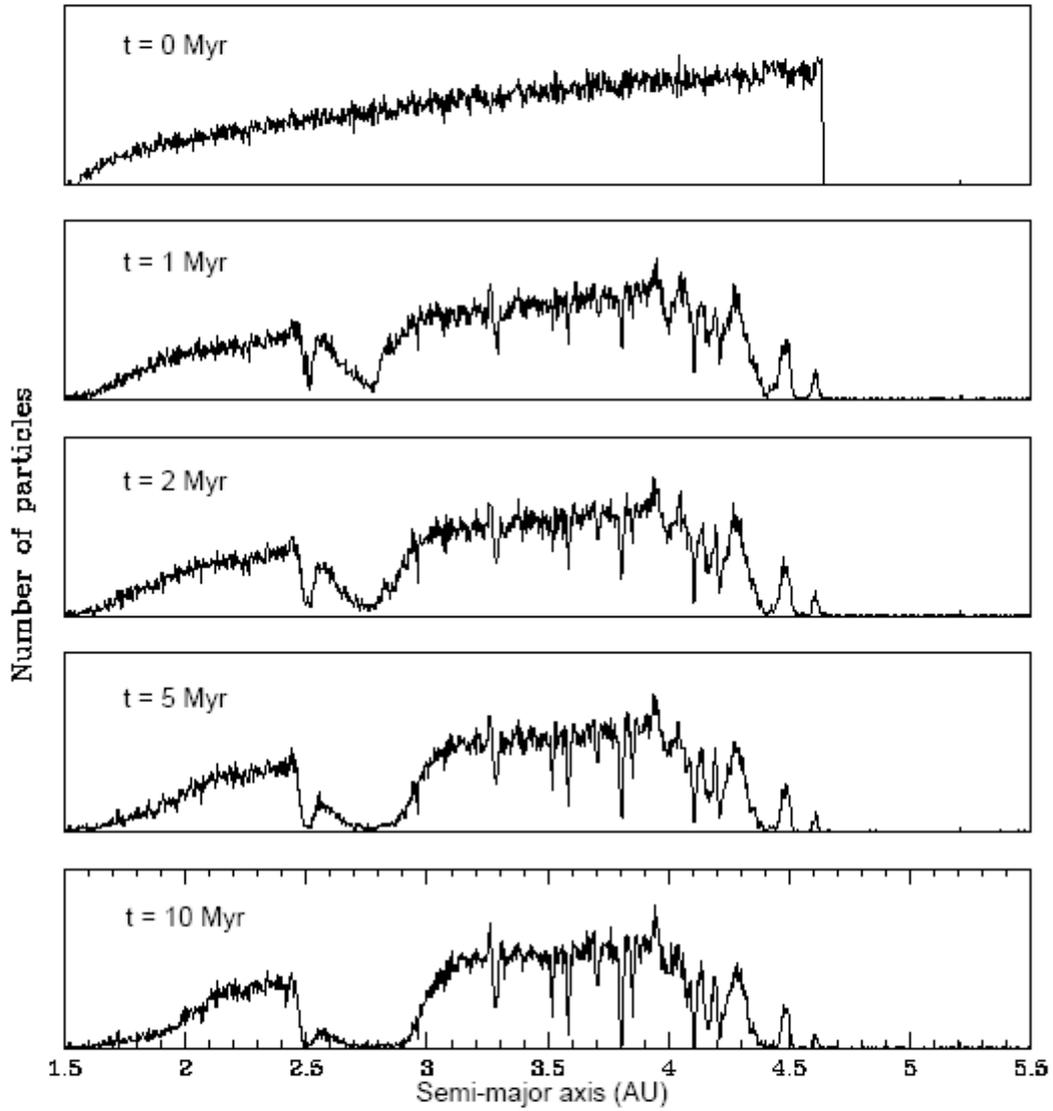

$M_{0.20}$

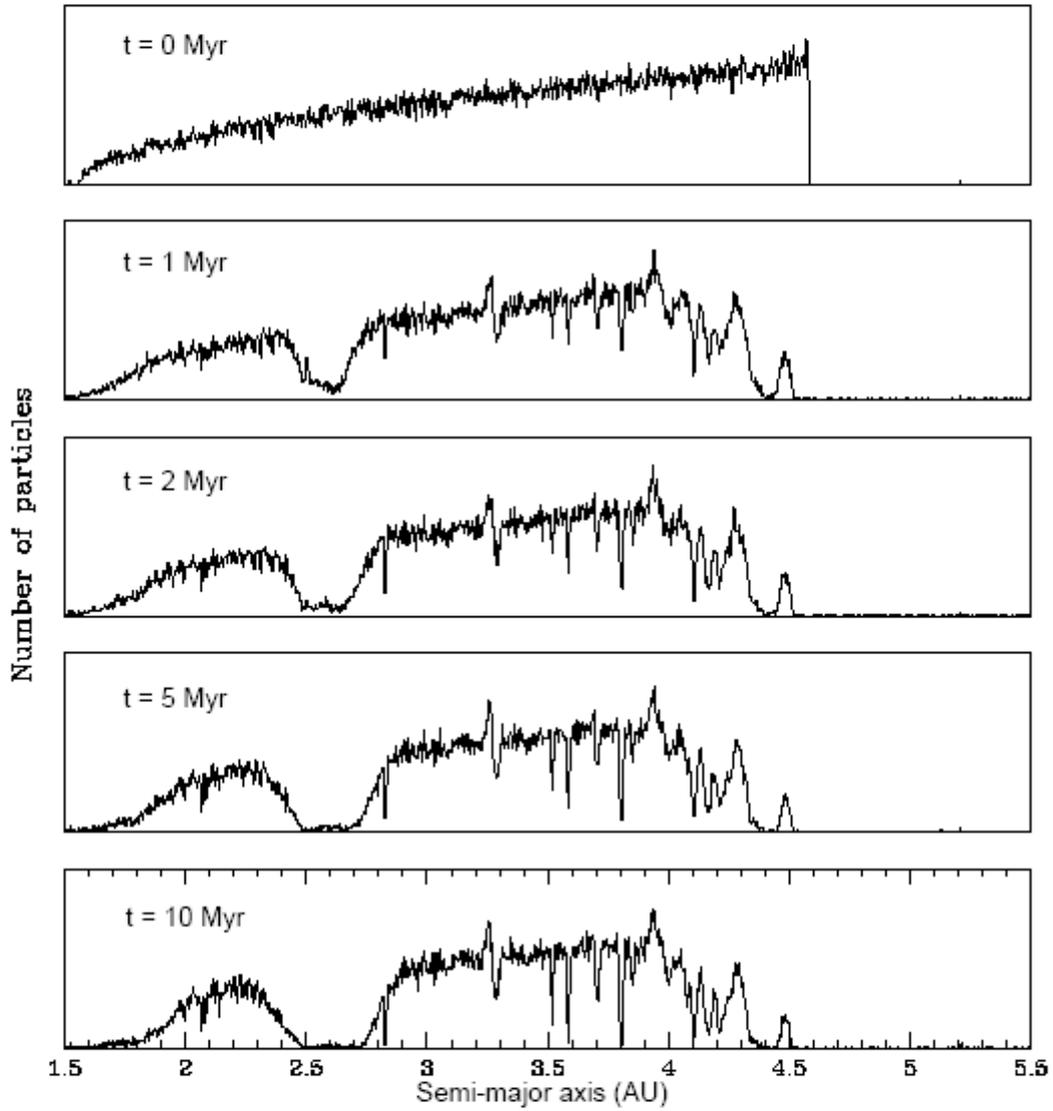

$M_{0.25}$

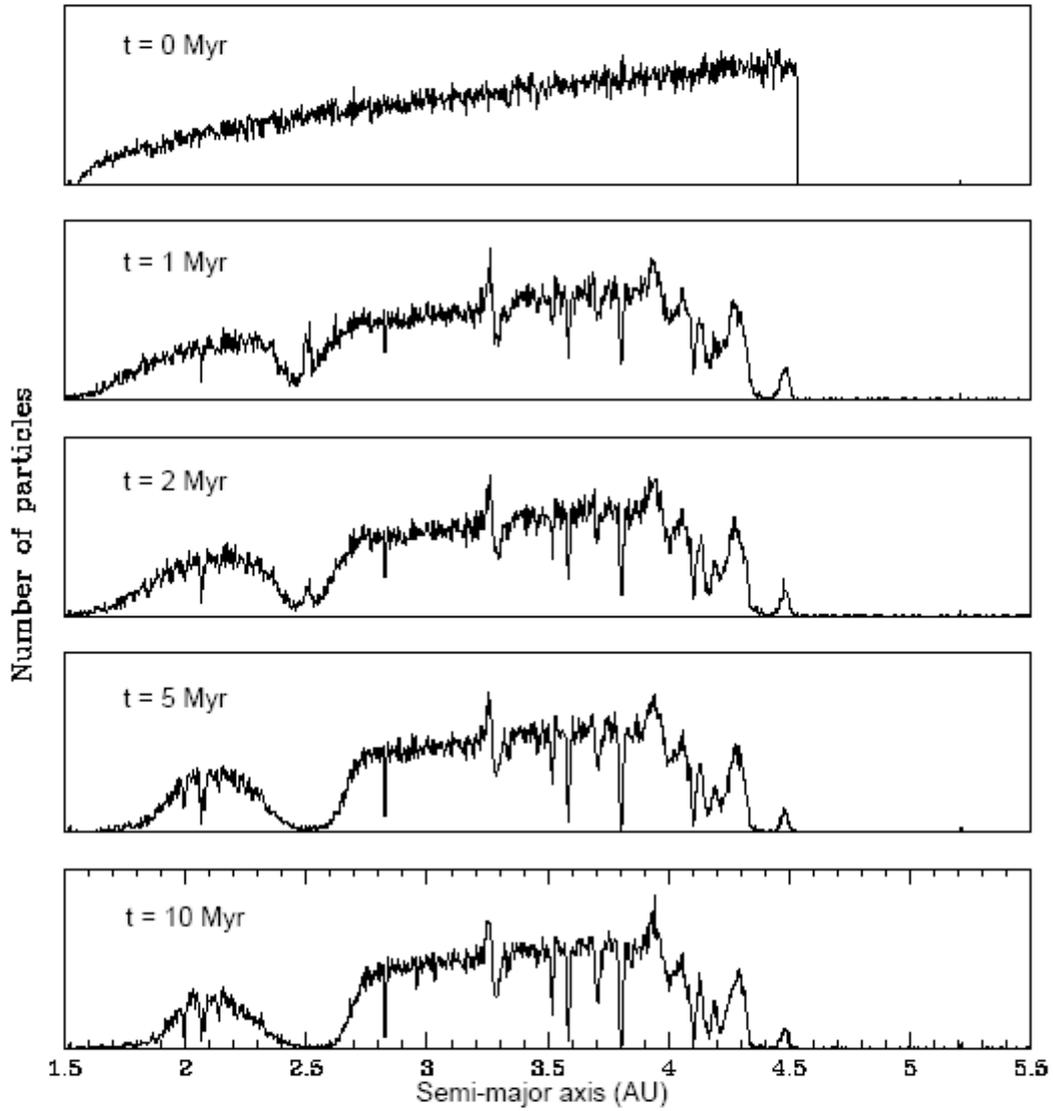

**M$_{0.33}$**

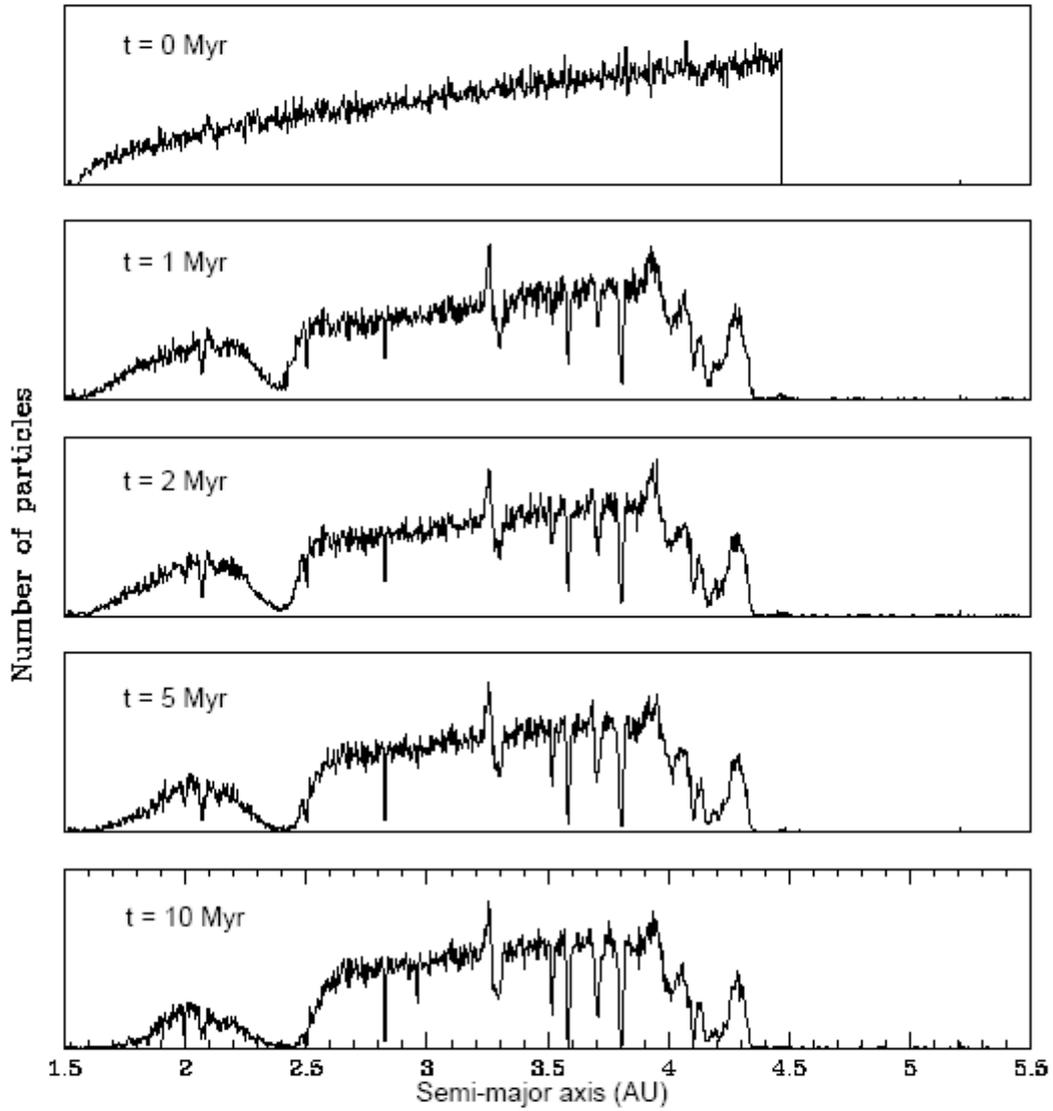

**M$_{0.50}$**

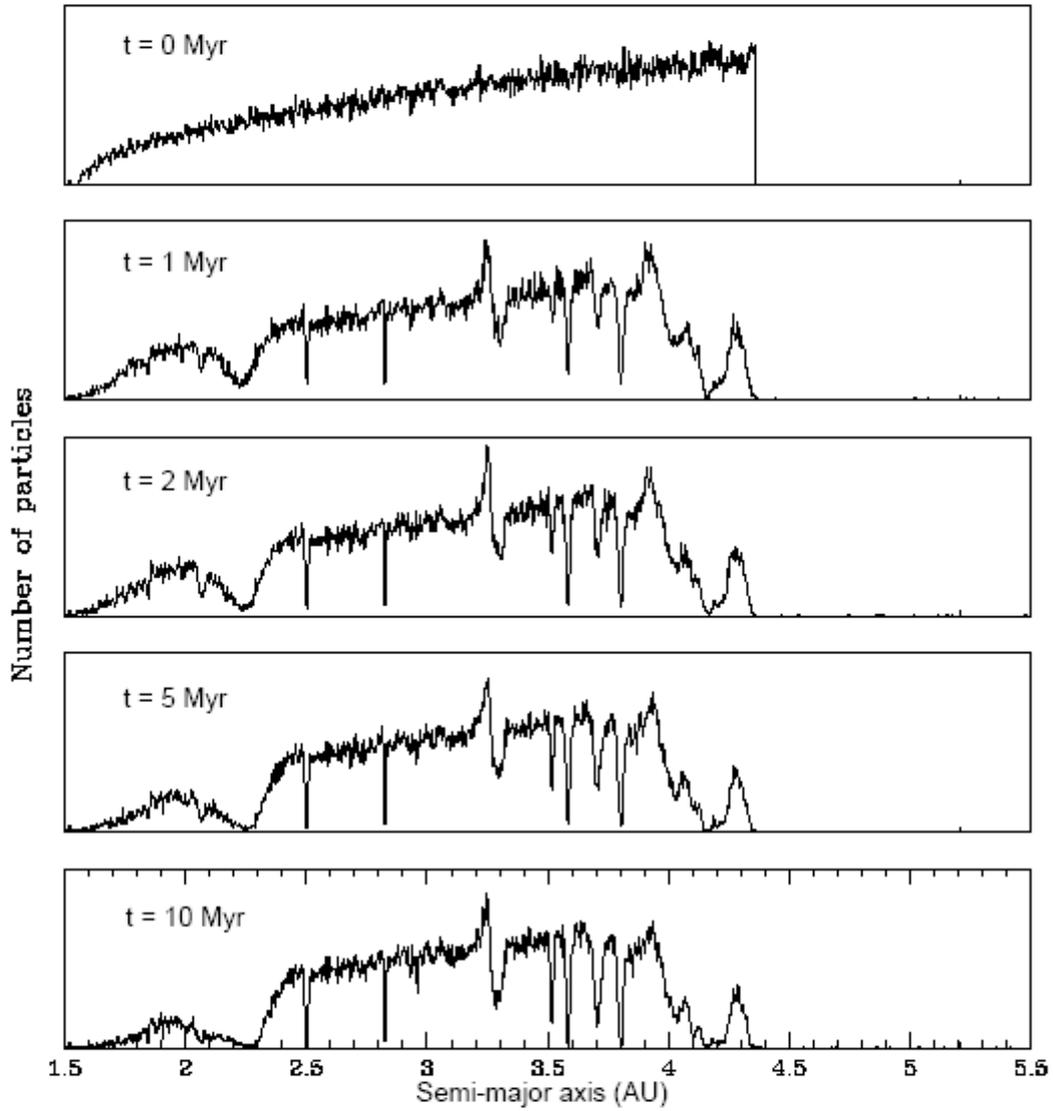

$M_{0.75}$

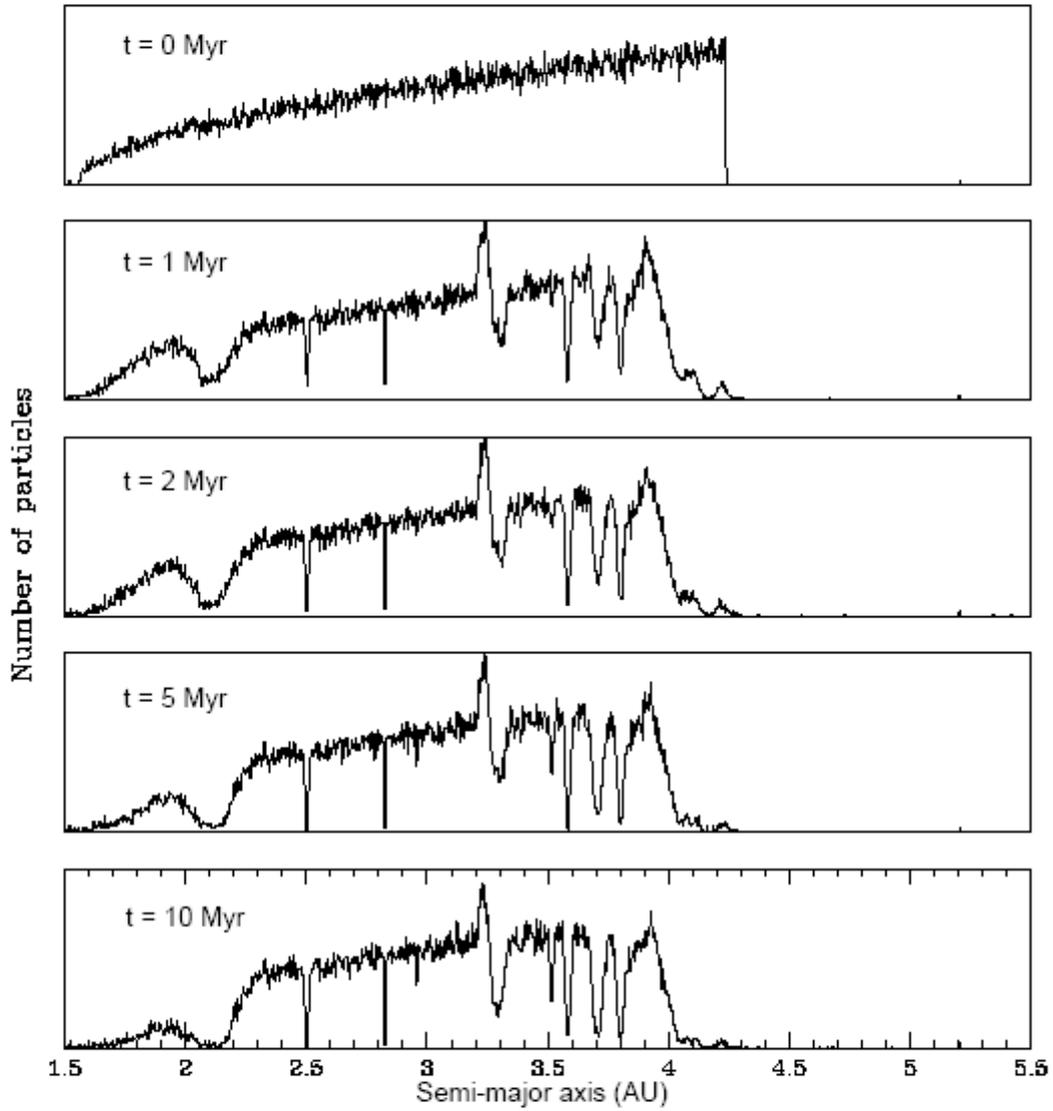

**M$_{1.00}$**

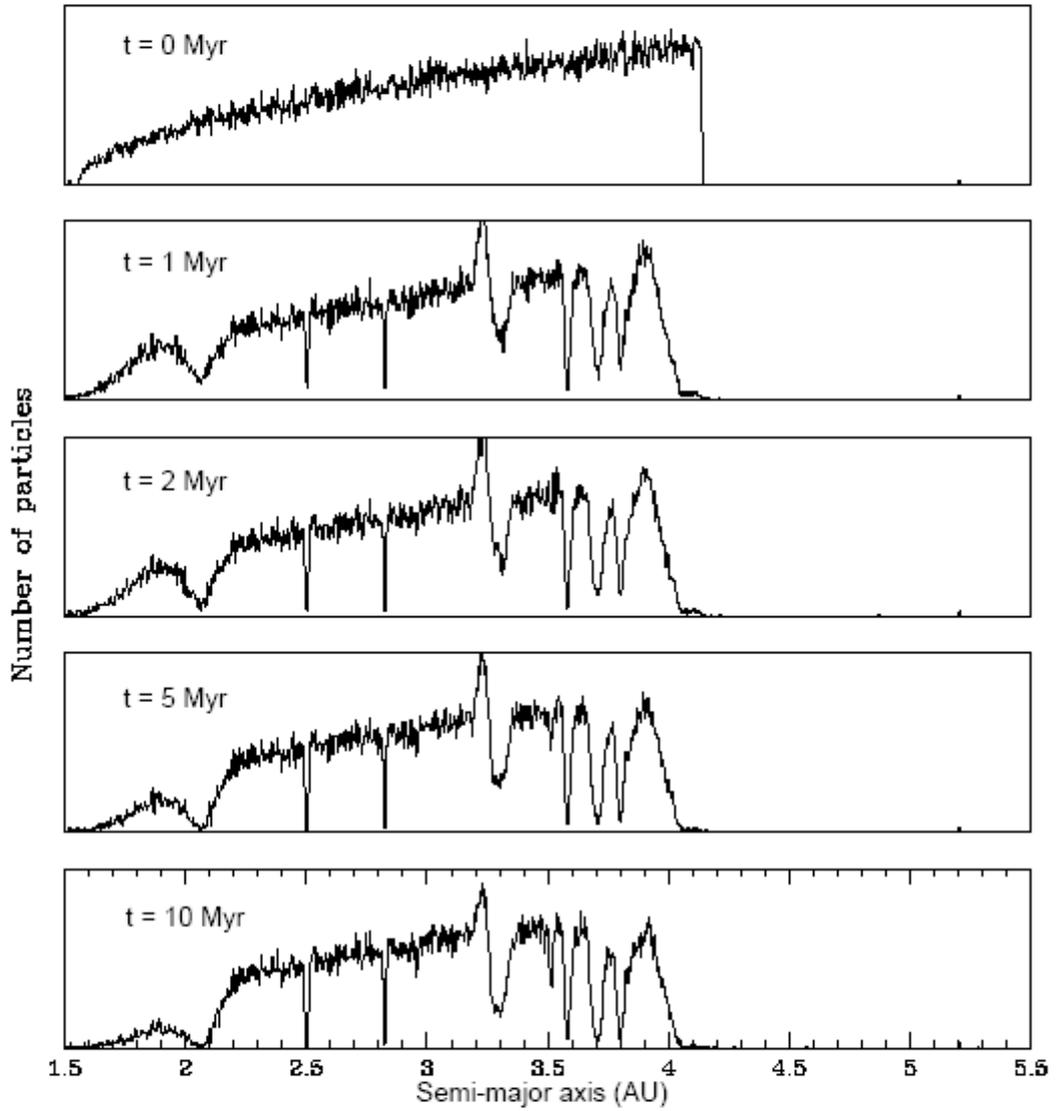

**M$_{1.50}$**

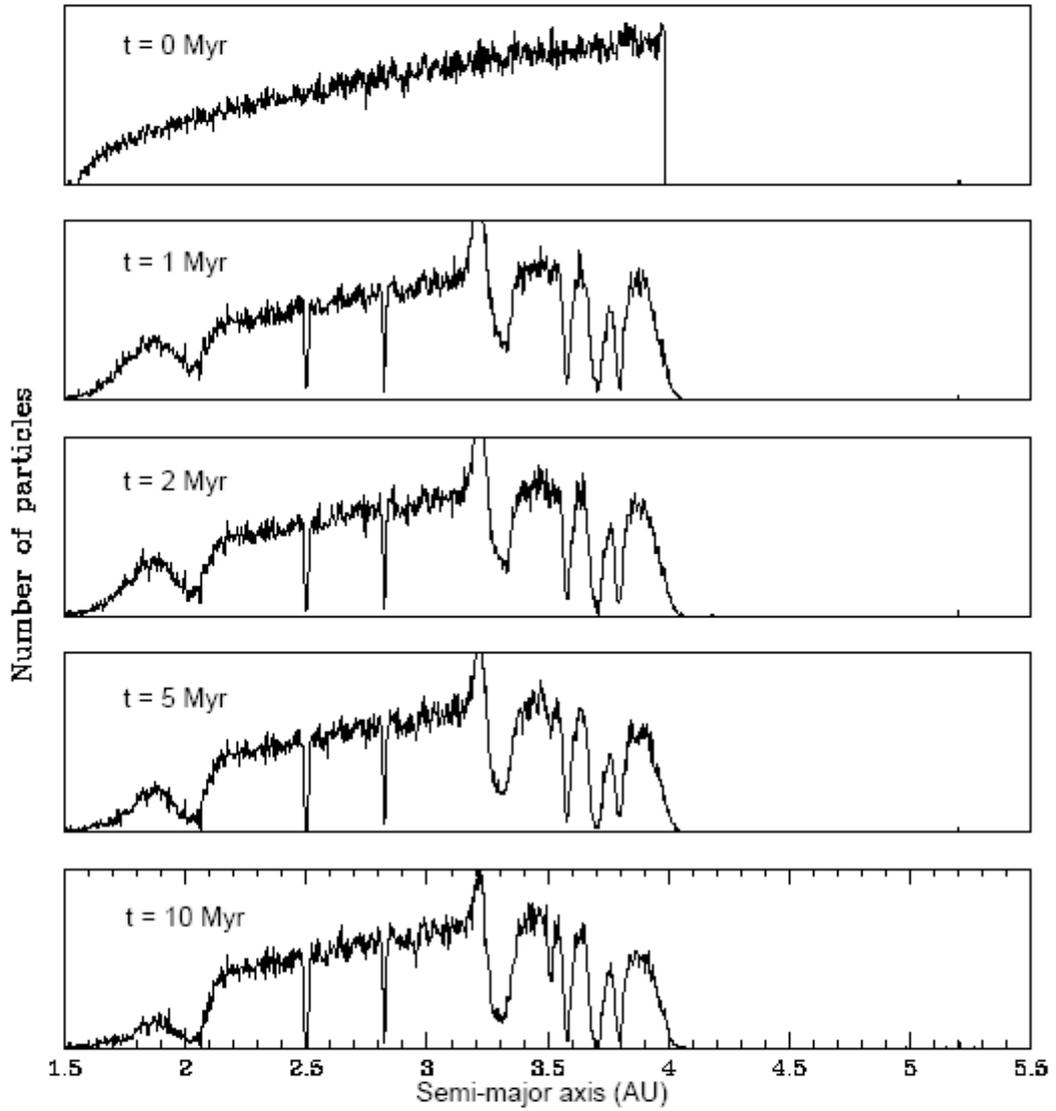

$M_{2.00}$

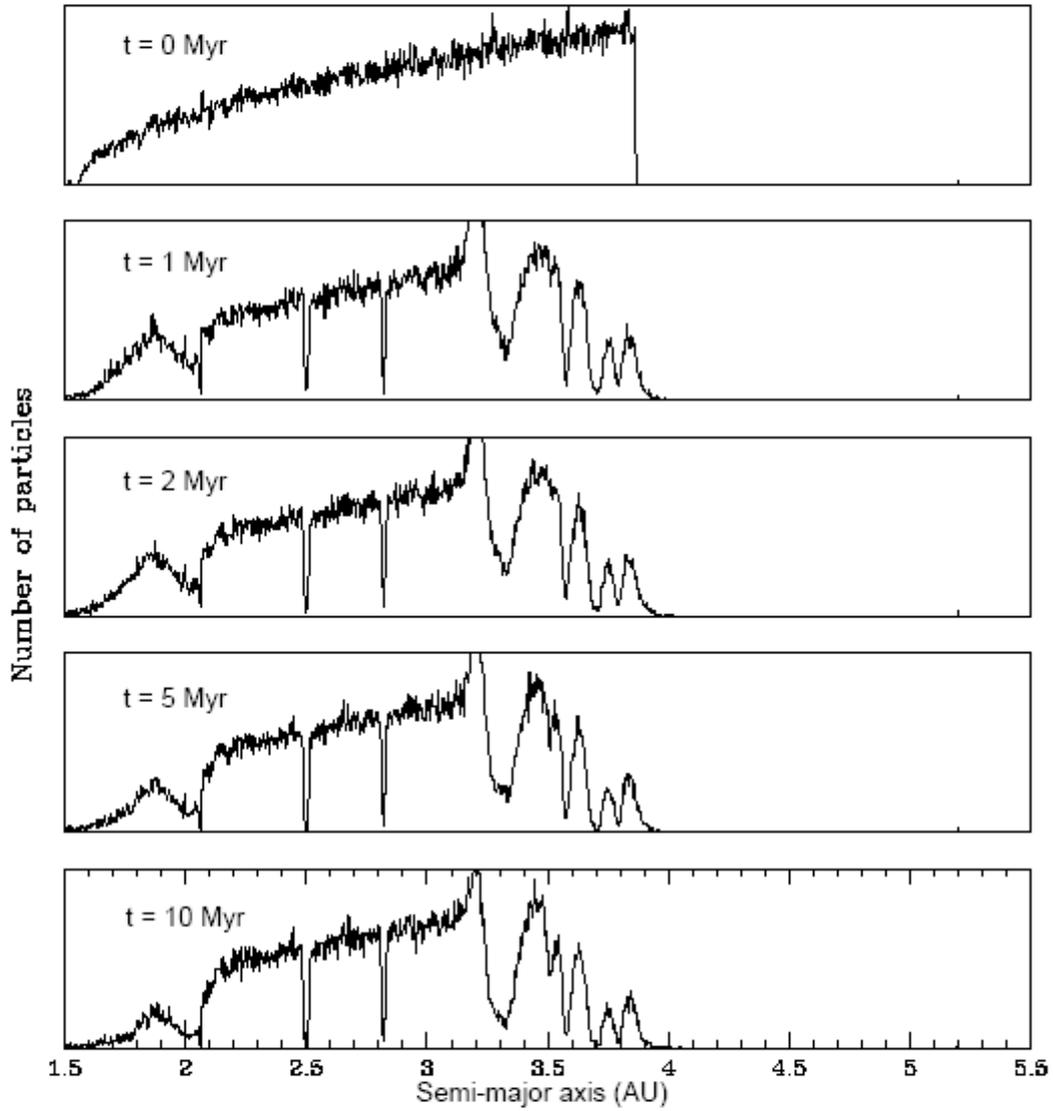

**Figures for inclusion with the text**

**Figure 1a – variation of population at $M_{0.25}$**

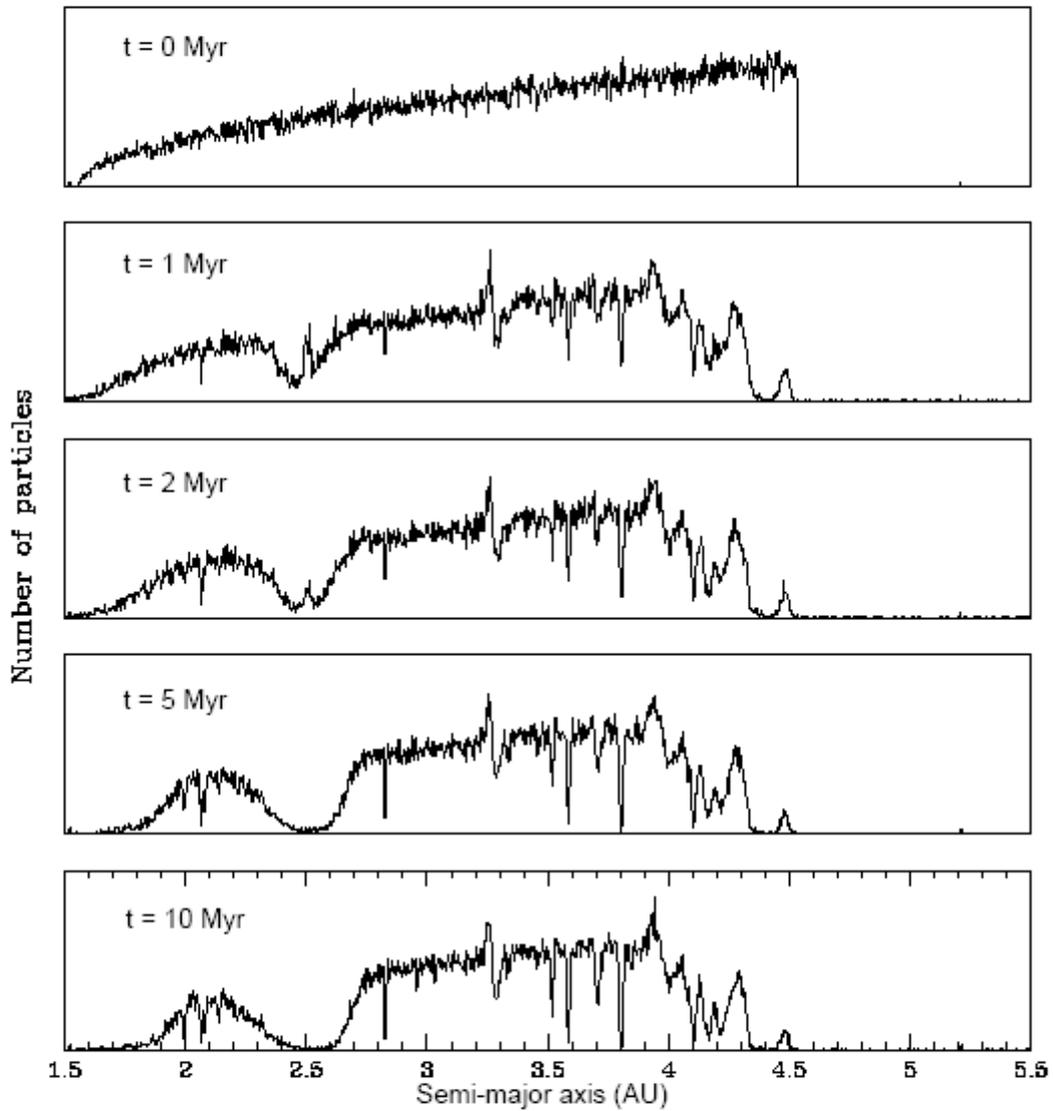

# Figure 1b. Variation of population at $M_{1.00}$

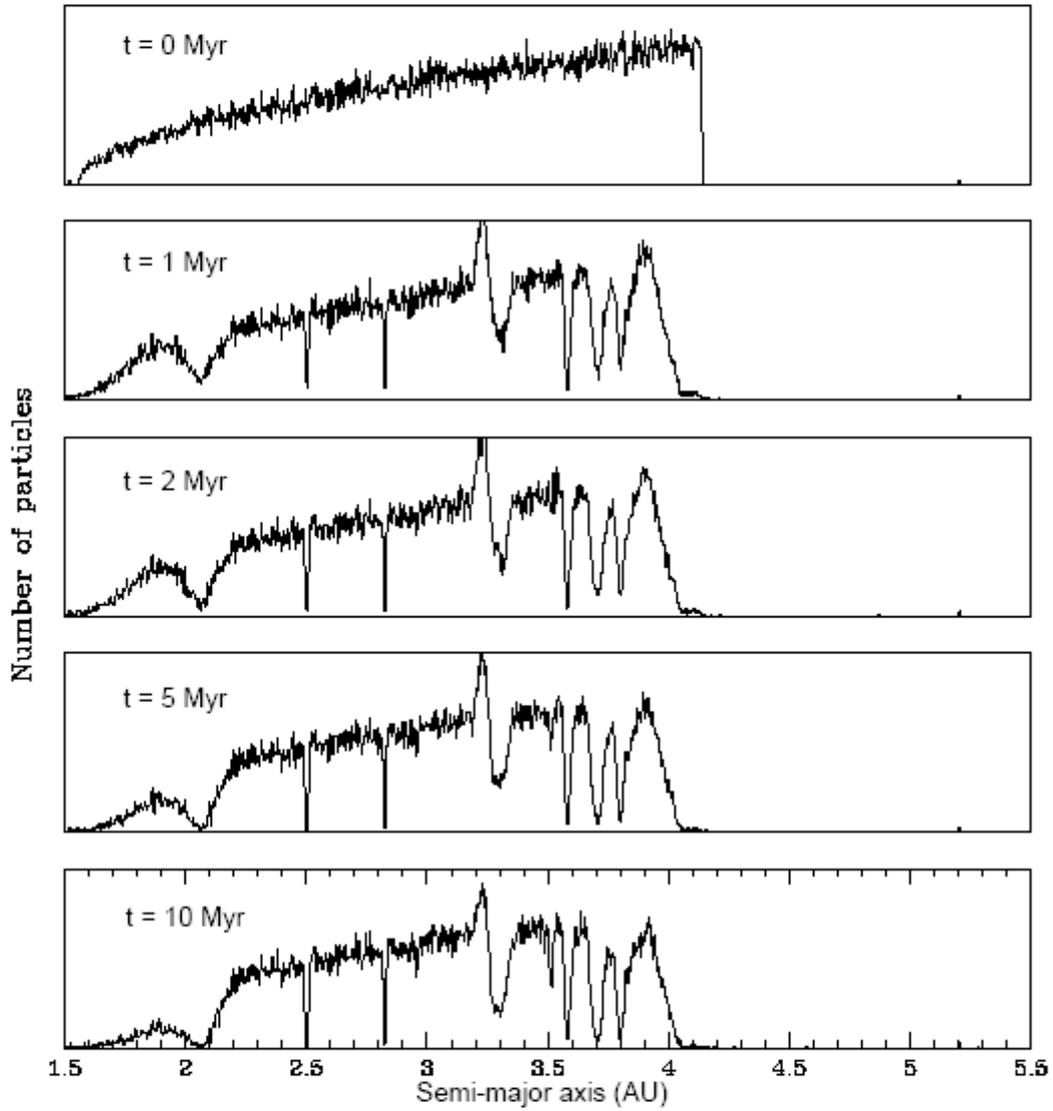

**Figure 2. Final distributions obtained at $M_{0.25}$ and $M_{1.00}$, showing the locations of key Mean-Motion Resonances.**

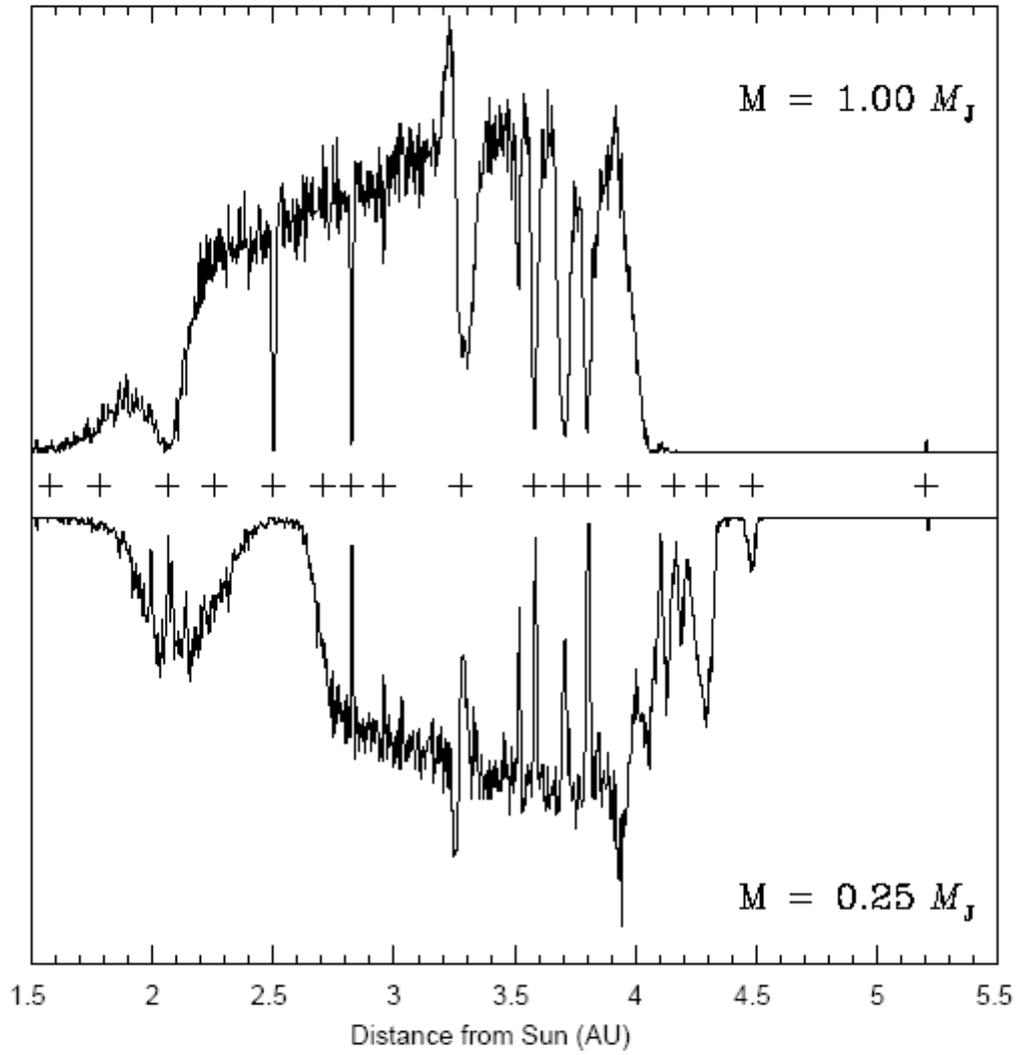

**Figure 3. Evolution of collision rate with Earth as a function of Jupiter mass**

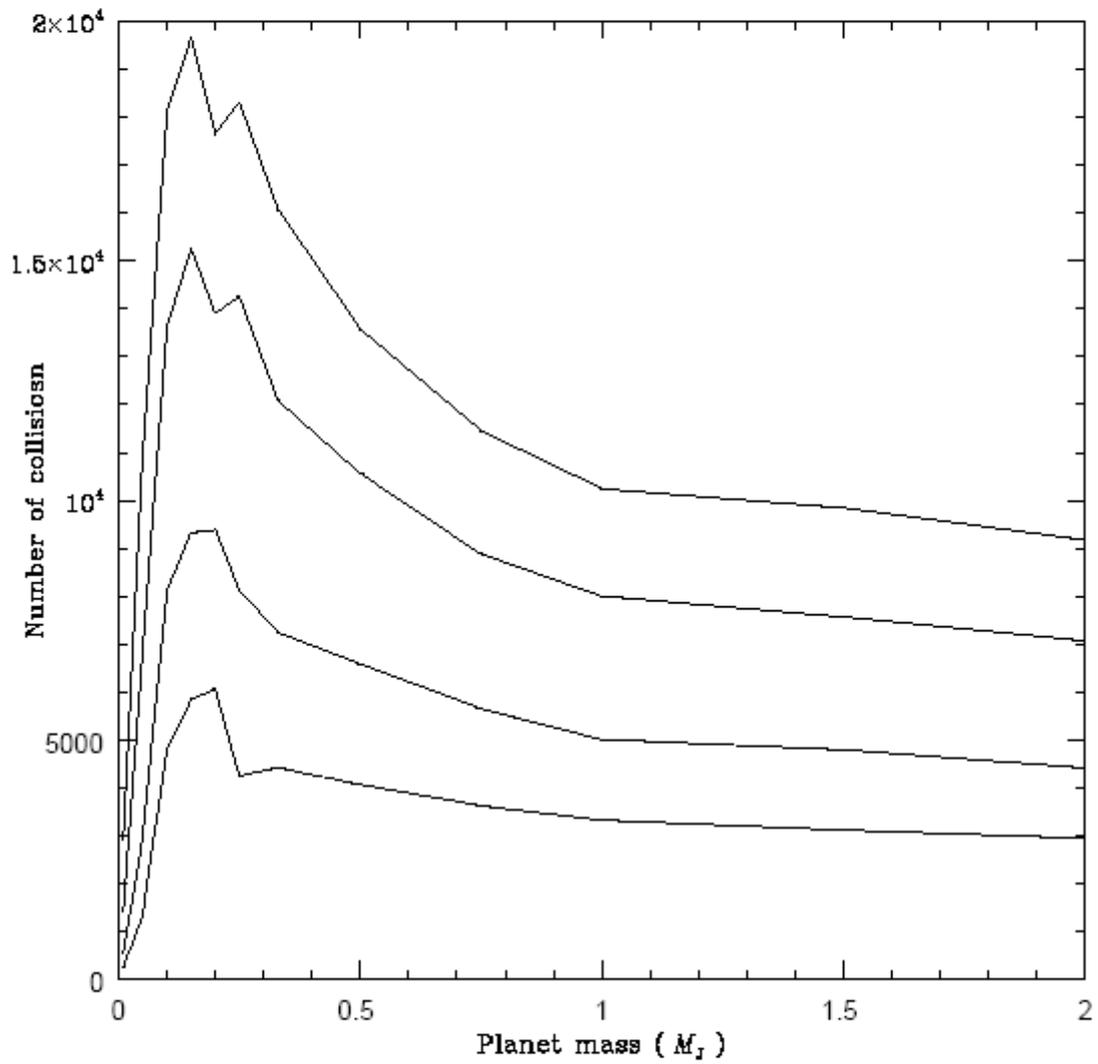

[i] In addition, the mass of Mars was increased slightly from its actual mass of 0.107 Earth masses, in order to account for any extra accretion which would have occurred as a result of a lower mass "Jupiter". The new Mars was given a somewhat arbitrary mass of 0.4 Earth masses. Rather than attempt to recursively modify the Mars mass as Jupiter itself varied, we chose a value intermediate between the current mass of the planet and that of the Earth. The mass of Mars makes little difference to our simulations, since it is held constant, and the planet is interior to the inner boundary of the belt. Even though a yet more massive Mars would have given a slightly larger perturbation to the inner asteroids, the small increase would have had no significant effect on our results.

[ii] Mean motion resonances are given in the form $n{:}m$, a simple integer ratio where, in the time it takes "Jupiter" to complete $n$ orbits another object completes $m$ orbits. For example, an asteroid located in the 3:7 mean motion resonance ($n = 3$, $m = 7$) would complete 7 orbits in the time it takes "Jupiter" to complete 3.

[iii] In much the same way as mean motion resonances result from a commensurability of the orbital periods of a planet and a given object, secular resonances occur as a result of commensurability between the precession rates of the perihelion or the longitude of the ascending node (or both). For example, if the ascending node of Jupiter's orbit precesses at the same rate as that of an asteroid, the two will be locked in a secular resonance, which can lead to significant alteration of the asteroids orbit over time, as energy is transferred between the two bodies. A detailed discussion of secular resonances is beyond the scope of this work, but we direct the interested reader to e.g. Murray and Dermott (1999(b)) for more information.